\documentclass[11pt]{article}
\pdfoutput=1
\usepackage{graphicx}
\let \ig=\includegraphics
\usepackage{color}
\begin{document}
\parindent0mm \unitlength1.0cm
\hspace{0cm}
\\[-1.2cm] 
\begin{center} 
{\Large \bf
The influence of the irregular forces on\\[0.2cm]   
the motions 
in the three-body problem}  
\\[0.7cm] 
Solovaya N.\,A., Sternberg Astronomical Institute,\\ Lomonosov Moscow State 
University,\\ University Prospect 13, 119\,899 Moscow, solov@sai.msu.ru
\end{center} 
\noindent
\\ 
{\bf Abstract} 
\\[0.3cm] 
The influence of the irregular forces on the evolution of a triple 
hierarchical stellar system  moving in the field of stars have been studied. 
Triple hierarchical stellar systems are stable in contrary to the stellar 
systems with comparable distances between all components. We considered the 
motion in the frame of the general tree-body problem using differential 
equations of the motion with the Hamiltonian without short-periodic terms.

\hspace*{0.5cm}
For isolated triple stellar systems, where we took into account the 
perturbations until the third order, we obtained the solution in which the 
mean motions of both components have the secular accelerations. Under the 
influence of perturbations of the distant component the mean motion in the 
near pair is slowed and vice versa. The mean motion of the distant star is 
constantly increasing. These changes are small, but on the cosmological time 
interval the hierarchical systems will convert into stellar systems, in which 
all components have comparable distances between each other. Such systems 
become unstable. 

\hspace*{0.5cm}
In a general case, if we take into account the irregular forces, the angular 
momentum of this system and its summary energy might be either loss or gain. 
These changes may influence on the dynamical evolution and stability of the 
stellar system.
\\
\par\noindent
{\bf Introduction} 
\\[0.3cm] 
We have investigated the dynamical evolution of the triple stellar system 
using the analytical theory of the particular case of the nonrestricted 
three-body problem (Orlov and Solovaya, 1988). Masses of components are 
comparable and the ratio of the semi-major axes of their orbits is the small 
parameter. The value of the eccentricities and the inclinations of both orbits 
are without any limitations. Unlike multiples systems with the comparable 
distances between components, these systems are thought to be stable. 

\hspace*{0.5cm}
The dynamical stability is understood as the conservation of the configuration 
of the system over the astronomical long time interval -- the eccentricities 
of the orbits remain less 1, the mutual inclinations change in small limits, 
there are no close approaches among the bodies, and the triple system is 
located inside of a stellar cluster. The study of the isolated triple stellar 
system, where we took into account the perturbations until the third order in 
the Hamiltonian (Solovaya, 1977), showed that the mean motions of both 
components have the secular accelerations (Solovaya, 2010).
 
\hspace*{0.5cm}
Due to the perturbations of the distant component the mean motion of the close 
pair slows, while for the distant component increases. These changes are 
small, but on the cosmological time interval hierarchical systems will 
converted into stellar systems, in which all components have comparable 
distances between each other.

\hspace*{0.5cm}
If the triple stellar system is located inside of a cluster, it must be 
subjected to perturbations of the surrounding stars. We have done  attempt to 
estimate the effect of their influence  on the dynamical evolution and 
stability of the triple stellar system, moving through the gravitational field 
of surrounding stars. In general case it may be either loss or gain of the 
kinetic energy. We used the Harrington $S$ criterion of stability   
(Harrington, 1972). For direct motion, when the mutual inclination of the two 
orbital planes is less than $90^\circ$,  $S = a_2\,(1-e_2)\,/\,a_1  > 3.5$. 
For the retrograde motion $S > 2.75$. Here $a_1$ and $a_2$ are the semi-major 
axes of the orbits and $e_2$  is the eccentricity of the distant component's 
orbit.
\\
\par\noindent
{\bf The gravitational perturbations } 
\\[0.3cm]
The motion of the isolated triple stellar system was studied using the  
analytical theory of the general three-body problem (Orlov and Solovaya, 1988).
We have used the Hamiltonian without short-periodic terms. As the intermediate 
orbits we used nonkeplerian ellipses. For the computation was used the 
Jacobian coordinate system and the canonical Delaunay elements -- $L_i$, 
$G_i$, $l_i$, $g_i$ (i=1,2) taking the invariable plane as a reference plane. 
Expanded in terms of the Legendre polynomials and truncated after the 
terms of the third order, the Hamiltonian has the form:
\begin{eqnarray}
F &\! \!=\!\! &  \frac{\gamma _{1}}{2L^{2}_{1}} + \frac{\gamma 
_{2}}{2L^{2}_{2}} -\frac{1}{16}
\gamma _{3}\frac{L_{1}^{4}}{L_{2}^{3}\,G_{2}^{3}}
\left[ \left( 1-3\,q^{2} \right) \left( 5-3\,\eta^{2} \right) -
\right.
\nonumber \\[0.2cm]  &&
\left.
- 15\,\left( 1-q^2 \right) \left( 1-\eta^2 \right)\,\cos{2\,g_1} \right]\,+R,
\end{eqnarray}
\\[-0.3cm]
where $R$ are the perturbations of the third order, the coefficients 
$\gamma_1$, $\gamma_2$, and $\gamma_3$ depend on masses, and 
\\[-0.5cm]
\begin{eqnarray}
q= \frac{c^{2}-G_{1}^{2}-G_{2}^{2}}{2 G_{1} G_{2}}, \ \ \ \ \ \
\eta=\sqrt{1-e_{1}^{2}}\,.
\end{eqnarray}
\\
$c$ is the constant of the angular momentum, $g_1$ is the argument of the 
periastron of the close pair in the invariable plane, and $q$ is the cosine of 
the mutual inclination of the orbits. The solution in hyperelliptic integrals 
$I_i$ $(i = 1,\,2,\,3)$  with the Hamiltonian until the terms of the second 
order was obtained by the method of Hamilton-Jacobi (Orlov and Solovaya, 1988).

\hspace*{0.5cm}
Denoting the mean motions of stars $\nu_i$, which equals $2\,\pi\,P_i^{-1}$,
where $P_i$ are the periods of revolution, we found  that the mean anomalies
$l_1$ and $l_2$ are expressed as: 
\begin{eqnarray}
l_1&=&B_1 +\nu_1\,\left(t-t_0\right)+
\,\,periodic\,\, terms,\nonumber \\[0.1cm]
l_2&=&B_2 +\nu_2\,\left(t-t_0\right)+
\,\,periodic\,\, terms\,.
\end{eqnarray}

a) The mean motion of the star $S_1$ 
\begin{eqnarray}
\nu_1&=&\left\{ 1+\frac{1}{16}\,\frac{\gamma\,m
^2}{\left( 1-e_2^2\right)^\frac{3}{2}} \left[ 4\,A_3+\frac{6\,
\delta}{\Sigma_1\,\overline{G}_2^2}\,
\left(\Sigma_1\,Q_1+\Sigma_2\,Q_2+
\,\Sigma_3\,Q_3\right)\right]\right\}\,n_1\,.
\end{eqnarray}
b) The mean motion of the star $S_2$ 
\begin{eqnarray}
\nu_2=\left[ m-\frac{1}{16}\,\frac{\gamma\,
^2}{\left( 1-e_2^2\right)^\frac{3}{2}}
\,\frac{3\,A_3\,\sqrt{1-e_2^2}}{\overline{G}_2}\right] n_1\,,
\end{eqnarray}
\\
where $\Sigma_i$ $(i=1,\,2,\,3)$ are the secular parts of the
hyperelliptic integrals.

\hspace*{0.5cm}
For stars moving along unperturbed orbits their mean motions may be 
expressed according to the third Keplerian law by formulae: 
\begin{eqnarray} 
n_1=k \sqrt{\frac{m_0+m_1}{a_1^{3}}}\,, \ \ \ \ \ \ \ \ \ \ \ \ \ \  
n_2=k \sqrt{\frac{m_0+m_1+m_2}{a_2{^3}}}\,,   
\end{eqnarray} 
where $a_1$ and $a_2$ are the semi-major axes of their orbits. Due to the 
mutual perturbations the value of the mean motion of the close pair slows by 
$\Delta n_1 = \nu_1-n_1$, while that of the distant component up by $\Delta 
n_2=\nu_2 -n_2$. 

\hspace*{0.5cm}
To the illustration of the theory the stellar system $\varepsilon$ Hydrae 
(ADS6993) was selected. The three brighter components of the system are moving 
on the orbits with the  semi-major axes $a_1=3.967$ AU and $a_2=75.76$ AU. The 
eccentricities of their orbits have values $e_1=0.67$ and $e_2=0.29$. The 
masses of the components in the solar mass are $m_0= 1.50$, $m_1= 1.30$, and 
$m_2=2.74$ (Heintz, 1963). 

\hspace*{0.5cm}
The results showed that for the close pair
\begin{eqnarray}
\nu_1 - n_1 =-0.019^{\circ} \,{\rm year}^{-1}\,,
\end{eqnarray} 
\\[-0.5cm]  
and for the distant component
\begin{eqnarray} 
\nu_2 - n_2 =+0.002^{\circ} \,{\rm year}^{-1}\,.
\end{eqnarray}
\\[-0.3cm] 
\hspace*{0.5cm}
Adding the perturbations of the third order $R$ in the Hamiltonian the mean 
motions $\nu_1$ and $\nu_2$ have the secular accelerations. The accelerations 
are small but on the cosmological time scale may  change the configuration of 
the system.

\hspace*{0.5cm}
The perturbation function has the following form:  
\begin{eqnarray}
R&=&\frac{15}{512}\,\gamma^4 \,\frac{L_1^6}{L_2^8}\, \frac{e_1\,e_2 
\sqrt{1-e_2^2}} 
{(1-e_2)^3 (1+e_2) ^3}\,\times \\[0.2cm]  \nonumber 
&&\times \left[ \left(4+3\,e_1^2\right)\left(-1-11\,q+5\,q^2+15 
\,q^3\right)\,\cos \left( g_1-g_2\right)\,\right] 
\end{eqnarray}
(Solovaya, 1997), where
\begin{eqnarray}
\gamma_4=k^2\,\mu_1 \,\frac{m_0-m_1}{m_0+m_1}\, \frac{\beta_1^6}{\beta_2^8}\,.
\end{eqnarray}

The mean motion $\nu_1$ of the close pair changes with time according to the 
formula
\begin{eqnarray}
\nu_1=n_{10} -\Delta\,n_1-\Delta\,n_1^{(3)} -\sigma_1\, t\,, 
\end{eqnarray} 
while the semi-major axis $a_1$ of the close pair increases.
Mean motion $\nu_2$ of the distant star increased  with time too, according to 
the formula
\begin{eqnarray} 
\nu_2=n_{20} +\Delta\,n_2+\Delta\,n_2^{(3)}  +\sigma_2\, t\,,
\end{eqnarray}
where $\sigma_1$ and $\sigma_2$ are the secular accelerations. 
Finally, the semi-major axis $a_2$ of the distant component decreases.

\hspace*{0.5cm}
The evolution of the semi-major axes $a_1$ and $a_2$ within the interval of 
$4.3\times 10^5$ years under the influence of the gravitational perturbation 
is shown in Figure 1 and 2. It is seen that under the influence of 
gravitational perturbations the triple hierarchical stellar system converts 
into the stellar system in which the components have comparable distance 
between each other. The coefficient $S$ is less than 3.5. Such system can be 
unstable.
\\
\par\noindent
{\bf Effect of the dynamical friction} 
\\[0.3cm]
Let the triple stellar system moves through the gravitational field of the 
surrounding stars in a stellar cluster. The triple system must be subjected to 
action of the irregular forces, due to random star approaches.  One of them is 
the dynamical friction. In general case, the triple stellar system  may loss 
or gain the kinetic energy.

\hspace*{0.5cm}
Stars in the cluster are considered as point masses interacting with each 
other according to the Newton's law. Each approach of the test star with the 
field star is based on the two-body approximation. Mass of the test star is 
bigger than mass of the field star. We have no information about the motion of 
the stars in the system. We attempt roughly to estimate the influence of 
dynamical friction on the motion of the stars in the triple system located in 
an isolated globular or open cluster. Such clusters contains several thousand 
stars. We shall assume that the velocities of the stars in the triple system 
not differ too much from the mean velocity of the field stars. Then the effect 
of the dynamical friction may be expressed by the following simplified 
equation: 
\begin{eqnarray}
\frac{\Delta v}{\Delta t}=-\eta v\,,
\end{eqnarray}
where $\eta$ is the coefficient of dynamical friction, $\Delta v$ is the mean 
change in the velocity of the test star during the time interval $\Delta t$.

\hspace*{0.5cm}
Chandrasekhar (1943) concluded that the coefficient of dynamical friction 
must be of the order of the reciprocal of the time of relaxation of the 
system. Consequently the effect of the dynamical friction can be apparent in 
systems with relatively short time of relaxation. We do not know the velocity
of the every star. But we can use the virial theorem, which confirms that for 
the stable state in the gravitating spherical distribution  of equal mass 
objects, the average potential energy must be equal the average kinetic 
energy, within a fractal two.

\hspace*{0.5cm}
From the virial theorem
\begin{eqnarray} 
2<K_{tot}> = <U> \,, \ \ \ \ \ \ \ \ \ \ 
<V^2> = G\,\frac{n\,m_f}{R_c}\,, 
\end{eqnarray}
where $<K_{tot}>$ is the average value of the kinetic energy and $<U>$ is the 
average value of the potential energy. $<V>$ is the average value of the 
velocity of a field star, $R_c$ is the radius of the cluster, $n$ is number 
of stars in the cluster, and $m_f$ is mass of the field star.

\hspace*{0.5cm}
Introduce the dynamical friction relaxation time $T_{fr}$ defined by the 
equation 
\begin{eqnarray} 
T_{fr} = \frac{2\,m_f}{m_f+m_t}\,T_d\,,   
\end{eqnarray} 
(Bertin, 2000), where $T_d$ is the relaxation time relative to the deflection, 
defined as 
\begin{eqnarray} 
T_d= \frac{N} {\ln (N)}\,\tau_d\,,
\end{eqnarray}
and $\tau_d$ is the natural crossing time
\begin{eqnarray} 
\tau_d = \frac{R_c}{<V>} \,.
\end{eqnarray}

Then the dynamical friction equation is 
\begin{eqnarray}
\frac{\Delta v}{\Delta t}  = - \frac{v_t}{T_{fr}}
= -\frac{4\,\pi\,G^2\,m_t\,\rho_f\,F\left( v \right) \,\ln{\Lambda}} 
{<v^3_t>}\,v_t\,, 
\end{eqnarray}
(Bertin, 2000), where $m_t$ is the mass of the test star, $m_f$ is the mass of 
the field star, $F(v)$ is the relation of the velocity $ v_f$ of the field 
star to the velocity $v_t$ of the test star, $\rho_f=m_f\,n$ is the mass 
density of the  cluster.
 
\hspace*{0.5cm}
The Coulomb logarithm is defined as
\begin{eqnarray}
\ln{\Lambda} = \ln{\frac{b_2} {b_1}}  \,,
\end{eqnarray}
where $b_1$ and $b_2$ are the minimum and maximum values of the impact 
parameters. 

\hspace*{0.5cm}
We used formulae of the two-body approximation. If the velocity of a star 
increases (decreases) on the $\Delta v $ in its apocenter (pericenter), then 
the  distance of the pericenter (apocenter) grows (falls) to the value  
\begin{eqnarray} 
4\,\Delta v\,\left[\frac{a^3\left( 1-e\right)} {\mu \left(1+e\right)} 
\right]^{\frac{1} {2}}\,,  
\end{eqnarray} 
where
$\mu=k^2\,\left( m_0+m_1\right)$. 
\newpage
\par\noindent
{\bf Results}
\\[0.3cm] 
We have used Equation (18) for the estimation of the change of  velocities 
of components in the triple stellar system in two cases, when this system is 
located in the globular cluster and in the open cluster. 

\hspace*{0.5cm}
It is known that the average linear diameter $R_c$ of the globular cluster 
range from 20 pc to 2000~pc. For our calculations we adopted values 
$n=10^6$, $R_c=35$ pc, and 
$G = 4.3\,10^{-3}$pc\,M$_\odot^{-1}$\,km$^2$\,s$^{-2}$.
For the globular cluster  with the equal star masses $m_f = 1$\,M$_\odot$ we 
see result in Figures 1 and 2. In this case the dynamical friction have no the 
influence on the evolution of the semi-major axis of the stars in the triple 
system.
\\[0.5cm] 
\par\noindent
\\
\begin{picture}(0,0) 
\put(0.0,-3.5){\ig[width=8.25cm,clip=]{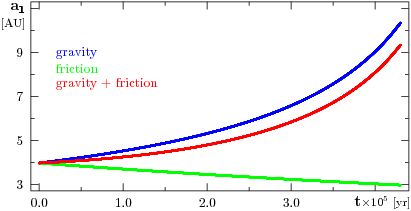}}
\put(8.75,-3.5){\ig[width=8.25cm,clip=]{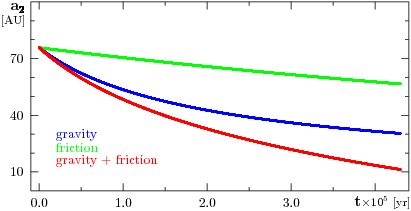}}
\end{picture}
\\[4.00cm] 
{\small {\bf Figure 1.}
\ \,The \,evolution \,of \,the \,semi-major \,axis \,of \,the \,triple 
\,stellar \,system \,in 
\,the \,globular \,cluster \hfill with\\$n=3\times 10^3$ members during the 
interval $ t = 4.3 \times 10^5$ years under the influence of gravitational 
perturbations, 
\centerline{dynamical friction and combined perturbations  for the growing 
values of $\Delta v$.}}     
\\[2.00cm]  
\begin{picture}(0,0) 
\put(0.0,-3.5){\ig[width=8.25cm,clip=]{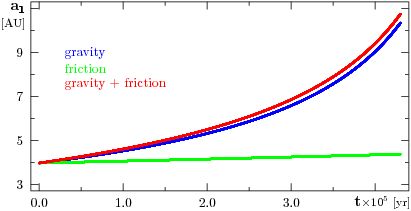}}
\put(8.75,-3.5){\ig[width=8.25cm,clip=]{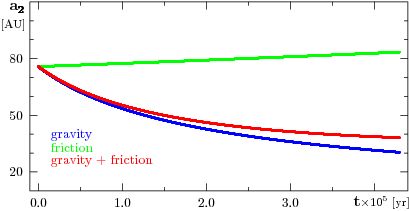}}
\end{picture}
\\[4.00cm] 
{\small {\bf Figure 2.}
The evolution of the semi-major axis of the triple stellar system  in the 
globular cluster with $n=3\times 10^3$ members during the interval $t = 4.3 
\times 10^5$ years under the influence of gravitational perturbations, 
\centerline{dynamical friction and combined perturbations  
for the slowing down values of $\Delta v$.}}      
\\[0.70cm] 
\par\noindent

\hspace*{0.5cm}
For the next example we took the open cluster similar to the Pleiades, with 
the number of members $n= 10^3$, the  mass $M = 800$\,M$_\odot$, and the 
radius of the cluster $R_c=2$\,pc. But  the impact parameter $b_1$ in Coulomb 
logarithm was taken as a free parameter. The values of the impact parameters 
were selected $b_1 =20$\,AU and $b_2 = 2$\,pc. The gravitational constant 
$G = 4.3\,10^{-3}$pc\,M$_\odot^{-1}$\,km$^2$\,s$^{-2}$.  

\hspace*{0.5cm}
The evolution of the semi-major axes are presented in Figures 3 and 4. It is 
seen that within the interval $t = 5.6 \times 10^5$ years the coefficient $S$ 
is greater than 3.5. If close approaches are possible, the dynamic friction 
has the influence on the evolution of the semi-major axis of the stars in 
the triple 
system.
\newpage
\par\noindent
\begin{picture}(0,0)  
\put(0.0,-4.0){\ig[width=8.25cm,clip=]{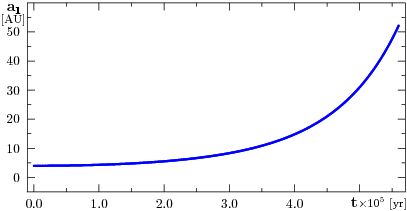}}
\put(8.75,-4.0){\ig[width=8.25cm,clip=]{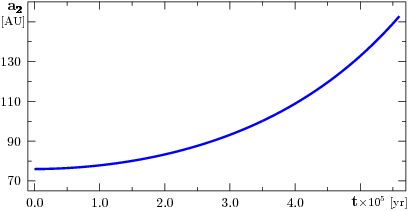}}
\end{picture}
\\[4.50cm] 
{\small {\bf Figure 3.}
The evolution of the semi-major axis of the triple stellar system in the open 
cluster with $n=3\times 10^3$ members during the interval $t = 5.6 \times 
10^5$ years under the influence of the dynamical friction for the growing 
\centerline{values of $\Delta v$.}}    
\\[1.60cm]  
\begin{picture}(0,0) 
\put(0.0,-3.5){\ig[width=8.25cm,clip=]{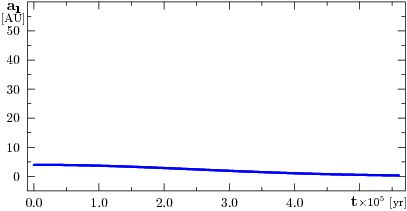}}
\put(8.75,-3.5){\ig[width=8.25cm,clip=]{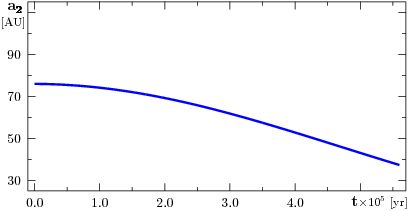}}
\end{picture}
\\[4.00cm] 
{\small {\bf Figure 4.}
The evolution of the semi-major axis of the triple stellar system in the open 
cluster with $n=3\times 10^3$ members during the interval $t = 5.6 \times 
10^5$ years under the influence of the dynamical friction for the slowing down 
\centerline{values of $\Delta v$.}}     
\\[0.6cm]  
{\bf Conclusions}
\\ 
\par\noindent
The mean motions of both components in the triple stellar system under the 
influence of the gravitational perturbations have the secular accelerations. 
The mean motion of the close pair is slowed and vice versa. The mean motion of 
the distant star is constantly increasing. On the cosmological time scale the 
configuration of the triple stellar system converts  from the hierarchical 
system into the stellar system, in which all components have comparable 
distances between each other. 

\hspace*{0.5cm}
The influence of the dynamical friction on the change of the semi-major axes 
of the stars in the triple stellar system is more less in the comparison with 
the gravitational perturbations of third order. In the case of a close 
approach of the field star to the triple stellar system the strong 
perturbations are possible. In this case the perturbations from the dynamical 
friction are greater than the gravitational perturbations of the third order. 
The influence of the dynamical friction on the evolution of the semi-major 
axes of the both orbits in the triple stellar system is evidently and the 
system can stay hierarchical. 
\\
\par\noindent
{\bf References} 
\\[-0.3cm]  
\par\noindent
Bertin, G.: 2000. {\it Dynamics of galaxies}. Cambridge Univ. Press, 74.
\\
Chandrasekhar, S.: 1943. {\it Astrophys. J.}, {\bf 97}, 255.
\\
Evans, D.: 1968. {\it Q. H. R. Astron. Soc.}, {\bf 9}, 388. 
\\
Harrington, R.: 1972. {\it Celestial Mech.}, {\bf 9}, 322.
\\
Heintz, W.\,D.: 1963. {\it Zeitschrift f\"ur Astrophysics}, {\bf 57}, 3.
\\
Orlov, A.\,A. and Solovaya, N.\,A.: 1988. In {\it Few Body Problem}. Ed. 
N.~Valtonen
Kluwer Acad. 
\hspace*{0.5cm}Publish., Dordrecht, 243.
\\
Solovaya, N.\,A.: 1977. {\it Vestn. Mosk. Univ. Fiz. Astron.}, {\bf 4}, 47.
\\
Solovaya, N.\,A.: 2010. {\it Vestn. Mosk. Univ. Fiz. Astron.}, {\bf 4}, 322.
\end{document}